\documentclass[12pt]{article}
\usepackage{times}
\usepackage{txfonts}
\usepackage{bm}
\usepackage{graphicx}
\usepackage{color}

\usepackage{natbib} 

\usepackage{sidecap} 
\sidecaptionvpos{figure}{t} 
\usepackage[font=footnotesize]{caption}

\usepackage[colorlinks=true, linkcolor=black, citecolor=black, urlcolor=blue]{hyperref}
\renewcommand\sec[1]{\vspace{0.05in}\noindent{{\large\bf{#1}}}
\addtocounter{section}{1}\setcounter{subsection}{0}
}
\newcommand\subsec[1]{\vspace{0.05in} \noindent{\bf{#1}}\addtocounter{subsection}{1}
}

\renewcommand{\b}{\begin{equation}}
\newcommand{\e}{\end{equation}}

\newcommand{\m}{\mbox{\scriptsize m}}
\newcommand{\rest}{\mbox{\scriptsize rest}}
\newcommand{\thMy}{{\mbox{\scriptsize th}}}
\newcommand{\reset}{{\mbox{\scriptsize reset}}}

\newcommand{\x}{\bm{x}}

\newcommand{\s}{ {\bm s}(t) }
\newcommand{\f}{ {\bm f}(t) }

\newcommand{\J}{\bm J}
\newcommand{\W}{{\bm W}}
\newcommand{\U}{{\bm U}}
\renewcommand{\u}{{\bm u}}

\newcommand{\V}{{\bm V}}

\newcommand{\Jtwit}{{\bm{\tilde J}}}
\newcommand{\utwit}{{\bm{\tilde u}}}
\newcommand{\Wtwit}{{\bm{\tilde W}}}
\newcommand{\Utwit}{{\bm{\tilde U}}}

\newcommand{\xMy}{ {\bm x}(t) }

\newcommand{\sMy}{\mbox{\scriptsize s}}
\newcommand{\fMy}{\mbox{\scriptsize f}}
\newcommand{\out}{\mbox{\scriptsize out}}
\newcommand{\inMy}{\mbox{\scriptsize in}}

\newcommand{\JMy}{\mbox{\scriptsize \sc J}}
\newcommand{\Fin}{F_{\inMy}(t)}
\newcommand{\Fout}{F_{\out}(t)}
\newcommand{\FJ}{{\bm F}_{\JMy}(t)}

\begin{document}

\thispagestyle{empty}
\vspace*{0.5in}
\begin{center}
\begin{Large}
{\bf
Using Firing-Rate Dynamics to Train\\ 
\vspace*{0.05in}
Recurrent Networks of Spiking Model Neurons\\
}
\end{Large}
\vspace*{0.2in}   
{\bf Brian DePasquale, Mark M. Churchland$^{1}$, L.F. Abbott$^{1,2}$}\\
\vspace*{0.1in}
Department of Neuroscience\\
$^1$Grossman Center for the Statistics of Mind\\
$^2$Department of Physiology and Cellular Biophysics\\
Columbia University College of Physicians and Surgeons\\
New York NY 10032-2695 USA\\

\vspace*{0.5in}
{\bf Abstract}
\end{center}

Recurrent neural networks are powerful tools for understanding and modeling computation and representation by populations of neurons. Continuous-variable or ``rate" model networks have been analyzed and applied extensively for these purposes.  However, neurons fire action potentials, and the discrete nature of spiking is an important feature of neural circuit dynamics.
Despite significant advances, training recurrently connected spiking neural networks remains a challenge.  We present a procedure for training recurrently connected spiking networks to generate dynamical patterns autonomously, to produce complex temporal outputs based on integrating network input, and to model physiological data. Our procedure makes use of a continuous-variable network to identify targets for training the inputs to the spiking model neurons. Surprisingly, we are able to construct spiking networks that duplicate tasks performed by continuous-variable networks with only a relatively minor expansion in the number of neurons. Our approach provides a novel view of the significance and appropriate use of ``firing rate" models, and it is a useful approach for building model spiking networks that can be used to address important questions about representation and computation in neural systems.
  
\newpage
\sec{Introduction}

A fundamental riddle of nervous system function is the disparity between our continuous and comparatively slow sensory percepts and motor actions and the neural representation of those percepts and actions by brief, discrete and spatially distributed actions potentials. A related puzzle is the reliability with which these signals are represented despite the variability of neural spiking across nominally identical performances of a behavior.  A useful approach to addressing these issues is to build spiking model networks that perform relevant tasks, but this has proven difficult to do.  Here we develop a method for constructing functioning networks of spiking model neurons that perform a variety of tasks while embodying the variable character of neuronal activity.  In this context, ``task" refers to a computation performed by a biological neural circuit.

There have been previous successes constructing spiking networks that perform specific tasks (see for example \citet{Seung, Wang, MachensBrody, Hennequin}).  In addition, more general procedures have been developed (reviewed in \citet{Abbottetal}) that construct spiking networks that duplicate systems of linear \citep{Eliasmith, BoerlinDeneve, BoerlinMachensDeneve} and nonlinear \citep{Eliasmith, Thalmeieretal} equations.  However, most tasks of interest to neuroscientists, such as action choices based on presented stimuli, are not expressed in terms of systems of differential equations.  

Our work  uses continuous-variable network models \citep{Sompolinsky}, typically called ``rate" networks, as an intermediary between conventional task descriptions in terms of stimuli and responses, and spiking network construction.  This results in a general procedure for constructing spiking networks that perform a wide variety of tasks of interest to neuroscience (see also \citet{Thalmeieretal, Abbottetal}).  We apply this procedure to example tasks and show how constraints on the sparseness and sign (Dale's law) of network connectivity can be imposed.  We also build a spiking network model that matches multiple features of data recorded from neurons in motor cortex and from arm muscles during a reaching task.

\sec{Results}

The focus of our work is the development of a procedure for constructing recurrently connected networks of spiking model.  We begin by describing the model-building procedure and then present examples of its use. 

\subsec{Network architecture and network training} \label{ssec:Network architecture}

\begin{figure} 
\centerline{ \includegraphics[width = 2in]{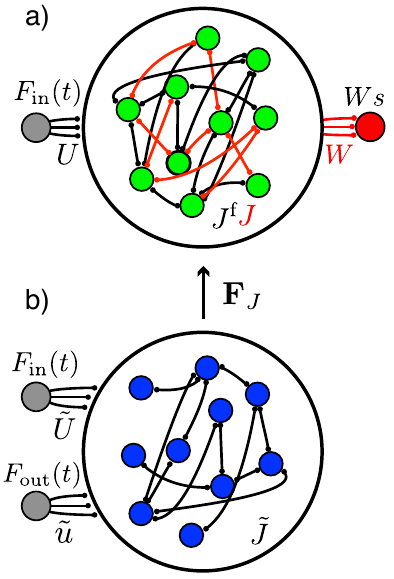}}
\caption{ Network architectures. {\bf a)} Spiking network.  A network of~$N$ recurrently connected leaky integrate-and-fire neurons (green circles) receives an input $\Fin$ (grey circle) through synapses~$\U$, and generates an output $\Fout$ (red circle) through synapses~$\W$. Connections marked in red (recurrent connections $\J$ and output connections $\W$) are modified by training, and black connections are random and remain fixed, including a second set of recurrent connections with strengths~$\J^{\fMy}$. {\bf b)} Continuous-variable network. A network of~$\tilde N$ recurrently connected ``rate" units (blue circles) receive inputs $\Fin$ and $\Fout$ through synapses~$\Utwit$ and $\utwit$, respectively.  All connections are random and held fixed. The sum of $\Utwit\Fout$ and the recurrent input determined by $\Jtwit$  defines the auxiliary targets $\FJ$ for the spiking network.}	
\label{fig:Figure1}
\end{figure}

The general architecture we consider is a recurrently connected network of~$N$ leaky integrate-and-fire  (LIF) model neurons that receives task-specific input~$\Fin$ and, following training, produces an approximation of a specified ``target" output signal~$\Fout$ (Figure \ref{fig:Figure1}a). $\Fin$ can be thought of as external sensory input or as input from another neural network, and $\Fout$ as the input current into a downstream neuron or as a more abstractly defined network output (for example a motor output signal or a decision variable). The neurons in the network are connected to each other by synapses with strengths denoted by the $N\times N$ matrix~$\J$.  Connections between the network and to the output have strengths given by an $N\times N_{\out}$ matrix~$\W$, where $N_{\out}$ is the number of outputs (either 1 or 2 in the examples we provide). During network training both~$\J$ and~$\W$ are modified. In addition to the trained connections described by $\J$, we also include random connections defined by another $N\times N$ matrix, $\J^{\fMy}$.  The elements of this matrix are chosen randomly from a Gaussian distribution with mean $\mu/N$ and variance $g_{\fMy}^2/N$ and are not modified during training. The values of $\mu$ and $g_{\fMy}$ are given below for the different examples we present. This random connectivity produces chaotic spiking in the network~\citep{VanVreeswijkSompolinsky, Brunel}, which we use as a source of spiking irregularity and trial-to-trial variability.  We use the parameter $g_{\fMy}$ to control the level of this variability.

When a neuron in the network fires an action potential, it contributes both fast and slow synaptic currents to other network neurons.  These currents are described by the two $N$-dimensional vectors, $\s$ and $\f$.  When neuron $i$ in the network fires an action potential, component $i$ of both $\s$ and $\f$ is incremented by 1, otherwise
\b
\tau_{\sMy}\frac{d\s}{dt} = -\s \quad\mbox{and}\quad \tau_{\fMy}\frac{d\f}{dt} = -\f \, .
\e
The two time constants determine the decay times of these slow and fast synaptic currents, and we set $\tau_{\sMy} = 100$ ms and  $\tau_{\fMy} = 2$ ms.  The synapses that are modified by training are all of the slow type, while the random synapses are fast.  For example, the output of the network is the product of~$\s$ and the output weight matrix~$\W$, $\W \s$ (Figure~\ref{fig:Figure1}a). 

The membrane potentials of the model neurons, collected together into a $N$-component vector $\V$, obey the equation  
\b
\tau_{\m}\frac{d\V}{dt} =  V_{\rest} - \V + g\Big(\J\s + \J^{\fMy}\f + \U\Fin\Big ) + I\, ,
\e
with $\tau_{\m} = 20$ ms.  For a case with $N_{\inMy}$ inputs, $\U$ is an $N\times N_{\inMy}$ matrix (we consider $N_{\inMy} = 1$ and 2) with elements drawn independently from a uniform distribution between -1 and 1\@.  $I$ is a bias current set equal to 10 mV\@.  It is increased between trials in the examples of Figures~\ref{fig:Figure3} and~\ref{fig:Figure4}, representing a ``holding" input.  Each neuron fires an action potential when its membrane potential reaches a threshold $V_{\thMy} = -55$ mV and is then reset to $V_{\reset} = V_{\rest} = -65$ mV\@.  Following an action potential, the membrane potential is held at the reset potential for a refractory period of 2 ms unless stated otherwise.  The parameter $g$ controls the strength of the inputs to each neuron, and we provide its value for the different examples below.

We can now specify the goal and associated challenges of network training. The goal is to modify the entries of~$\J$ and~$\W$ so that the network performs the task specified by~$\Fin$ and~$\Fout$, meaning that
\b
	\label{eq:ZWsFout}
	\W \s \approx \Fout 
\e
when the network responds to~$\Fin$ (with the approximation being as accurate as possible). \hyperref[eq:ZWsFout]{Equation~\ref{eq:ZWsFout}} stipulates that $\s$ must provide a basis for the function~$\Fout$.  If it does, it is straightforward to compute the optimal $\W$ by minimizing the squared difference between the two sides of equation~\ref{eq:ZWsFout}, averaged over time.  This can be done either recursively \citep{Haykin} or using a standard batch least-squares approach.

Determining the optimal $\J$ is significantly more challenging because of the recurrent nature of the network.  $\J$ must be chosen so that the input to the network neurons, $\J\s$, generates a pattern of spiking that produces $\s$.  The circularity of this constraint is what makes recurrent network learning difficult.  The difference between the easy problem of computing $\W$ and the difficult problem of computing $\J$ is that, in the case of $\W$, we have the target $\Fout$ in equation~\ref{eq:ZWsFout} that specifies what signal $\W$ should produce.  For $\J$, it is not obvious what the input it generates should be.  

Suppose that we \emph{did} have targets analogous to $\Fout$ but for computing $\J$ (we call them auxiliary target functions and denote them by the~$N$-component vector~$\FJ$).  Then, $\J$, like $\W$, could be determined by a least-squares procedure, that is, by minimizing the time-averaged squared differences between the two sides of
\b
	\label{eq:zJsf}
	\J \s \approx \FJ \, .
\e
There are stability issues associated with this procedure, that we discuss below, however the main challenge in this approach is to determine the appropriate auxiliary target functions. Our solution to this problem is to obtain them from a continuous-variable model.  More generally, if we can train or otherwise identify another model that implements a solution to a task, we can use signals generated from that model to train our spiking network.

\subsec{Using continuous variable models to determine auxiliary target functions}

Equations~\ref{eq:zJsf} and~\ref{eq:ZWsFout}, respectively, summarize two key features of the vector of functions $\FJ$: 1) They should correspond to the inputs of a recurrently connected dynamic system, and 2) They should provide a basis for the network output $\Fout$.  To satisfy the first of these requirements, we identify $\FJ$ with the inputs of a recurrently connected continuous-variable ``rate" network. These networks have been studied intensely~\citep{Sompolinsky, RajanAbbottSompolinsky} and have been trained to perform a variety of tasks \citep{Jaeger, SussilloAbbott, Buonomano, SussilloRev}.  To satisfy the second condition, we use the desired spiking network output, $\Fout$, as an \emph{input} to the rate network.  This allows us to obtain the auxiliary target functions without having to train the continuous variable network. 

The continuous-variable model we use is a randomly connected network of~$\tilde N$ firing-rate units (throughout we use tildes to denote quantities associated with the continuous-variable network).  Like the spiking networks, these units receive the input~$\Fin$ and, as mentioned above, they also receive $\Fout$ as an input. The continuous-variable model is described by an $\tilde N$-component vector $\xMy$ that satisfies the equation
\b
\tau_x\frac{d\xMy}{dt} = -\xMy + \tilde g\Jtwit H(\xMy) + \utwit\Fout + \Utwit\Fin \, ,
\e
where $\tau_x = 10$ ms, $H$ is a nonlinear function (we use $H(\cdot) = \tanh(\cdot)$), and $\tilde J$, $\utwit$, and $\Utwit$ are matrices of dimension $\tilde N\times\tilde N$, $\tilde N\times N_{\out}$ and  $\tilde N\times N_{\inMy}$, respectively.  The elements of these matrices are chosen independently from a Gaussian distribution of zero mean and variance $1/N$ for $J$, and a uniform distribution between -1 and 1 for $\utwit$ and $\Utwit$ unless stated otherwise.  We set $\tilde g = 1.2$ except where identified otherwise.

To be sure that signals from this driven network are appropriate for training the spiking model, the continuous-variable network, driven by the target output, should be capable of producing a good approximation of $\Fout$.  To check this, we can test whether an $N_{\out}\times \tilde N$ matrix can be found (by least squares) that satisfies $\Wtwit H(\xMy) \approx \Fout$ to a sufficient degree of accuracy. Provided~$\Jtwit$ and~$\utwit$ are appropriately scaled, this can be done for a wide range of tasks \citep{SussilloRev}. 

The auxiliary target functions~$\FJ$ that we seek are generated from the inputs to the neurons in the continuous-variable network.  There is often, however, a mismatch between the dimensions of~$\FJ$, which is $N$, and of the inputs to the continuous-variable model, which is $\tilde N$.  To deal with this, we introduce an $N\times\tilde N$ matrix $\u$, with elements drawn independently from a uniform distribution over the range $\pm\sqrt{3/\tilde N}$, and write
\b
	\label{eq:Jsu}
	\FJ = \u\Big(\tilde g\Jtwit H(\xMy) + \utwit \Fout\Big) \, .
\e
We leave out the input term proportion to $\Fin$ in this expression because the spiking network receives the input $\Fin$ directly.  This set of target functions satisfies both of the requirements listed at the beginning of this section and, as we show in the following examples, they allow functional spiking networks to be constructed by finding connections $\J$ that satisfy equation~\ref{eq:zJsf}.  We do this initially by a recursive least squares algorithm \citep{Haykin}, but later we discuss solving this problem by batch least squares instead.

\subsec{Examples of trained networks}

The procedure described above can be used to construct networks that perform a variety of tasks. We present three examples that range from tasks inspired by problems of relevance to neuroscience to modeling experimental data. 

\begin{figure}
\centerline{\includegraphics[width=2in]{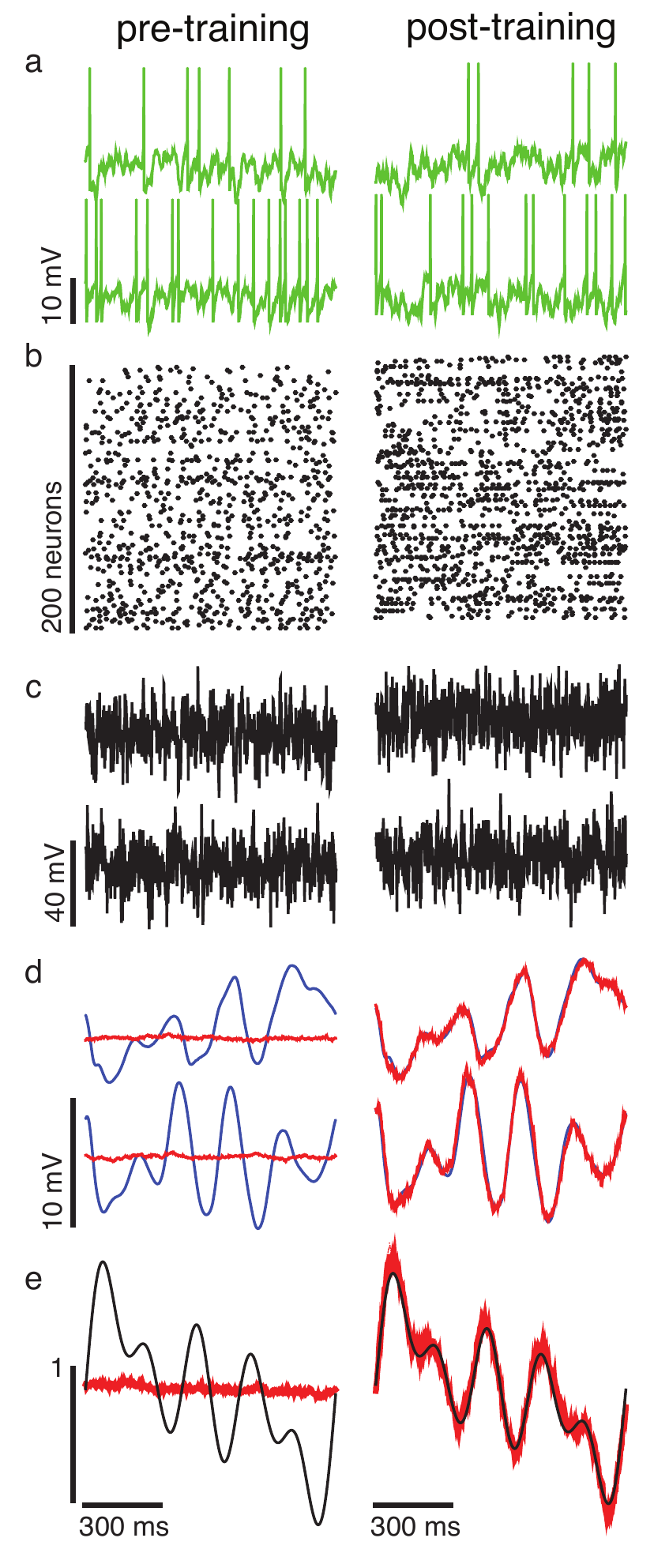}}
	\caption{Network activity before and after training for an autonomous oscillation task. {\bf (a)} Membrane voltage and spiking activity of two example neurons. {\bf (b)} Raster plot of 200 neurons. {\bf (c)} Random recurrent input~$\J^{\fMy}\f$ for two example neurons.  {\bf (d)} Auxiliary target function~$\FJ$ (black) and learned recurrent input~$\J\s$ (red) for two example neurons. {\bf (e)} Target output~$\Fout$ (black) and the generated output~$\W\s$ (red) over one period.  Parameter for this example: $N = 3000$, $g = 7$ mV, $\mu = -57$, $g_{\fMy} = 17$, and $\tilde N = 1000$.}
\label{fig:Figure2}	
\end{figure}

\begin{figure}
\centerline{\includegraphics[width=3in]{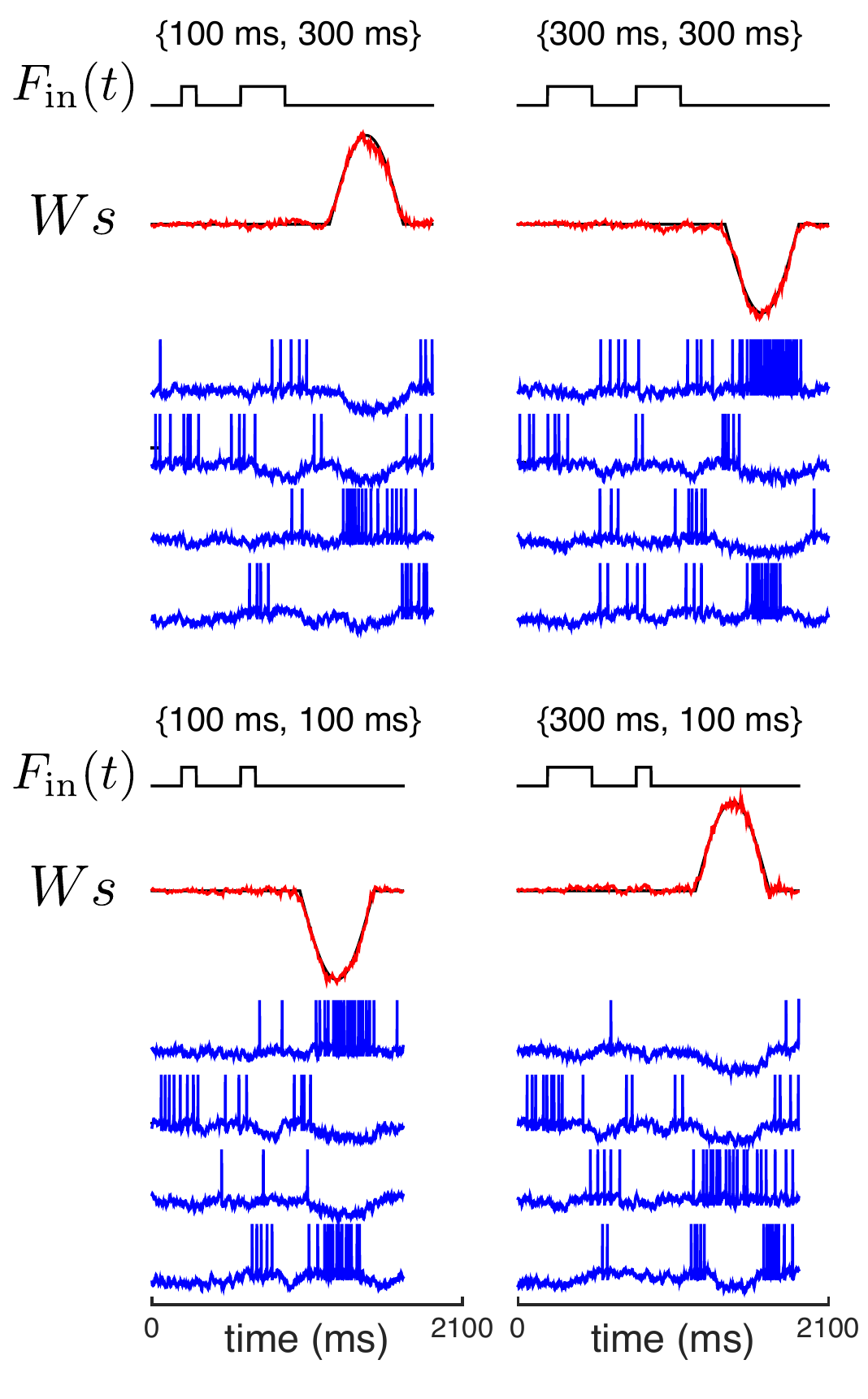}}
\caption{Temporal XOR task.  The input $\Fin$ (black) consists of two pulses that are either short or long in duration. The output~$\W\s$ (red) should report an XOR function of the combination of pulses.  Membrane potentials of 10 example neurons (blue) are shown for the 4 different task conditions. Parameters for this example: $N = 3000$, $g = 10$ mV, $\mu = -40$, $g_{\fMy} = 12$ and $\tilde N = 1000$.}
\label{fig:Figure3}	
\end{figure}

\begin{figure}
\centerline{\includegraphics[width=3.5in]{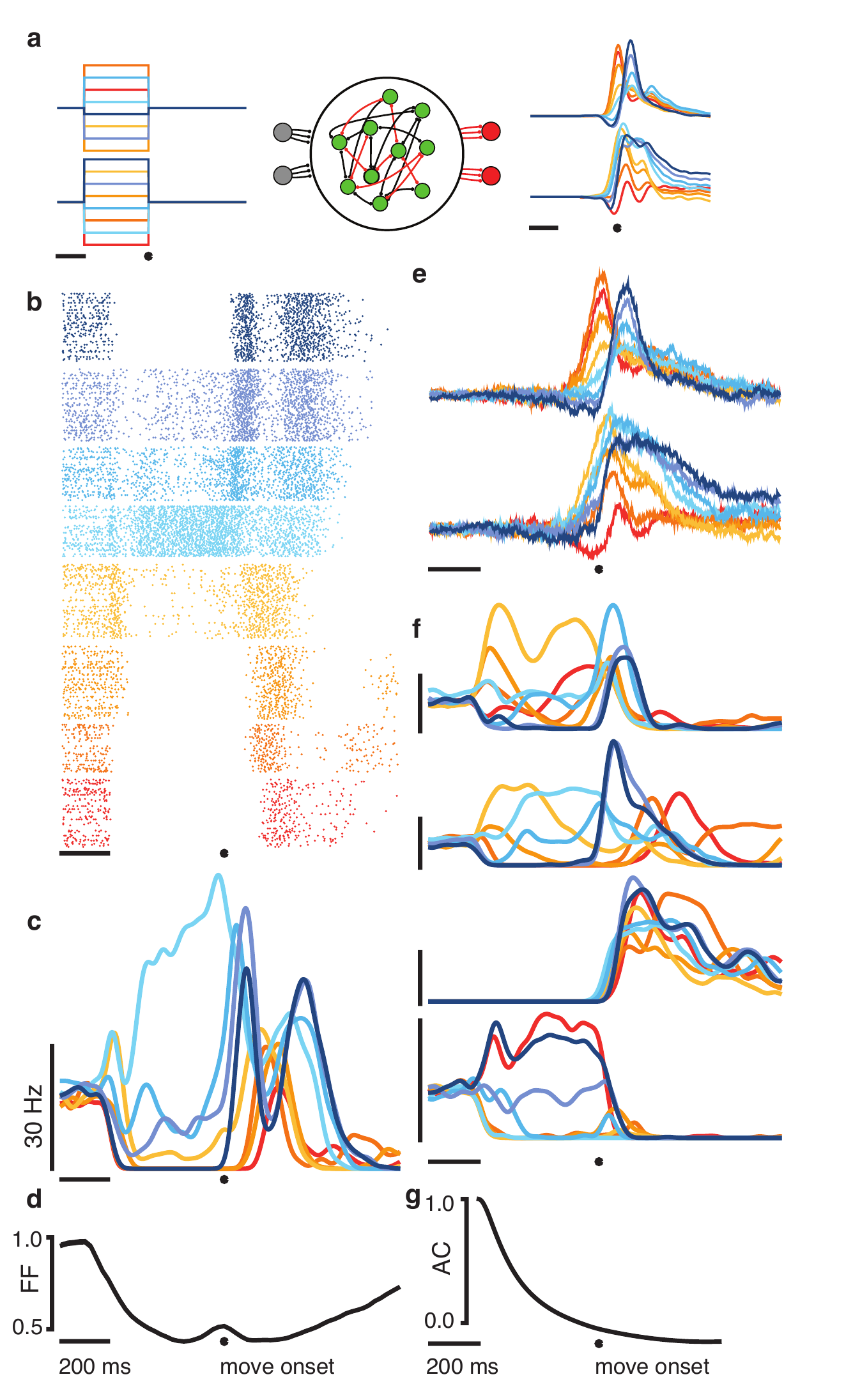}}
\caption{Producing EMG during reaching. {\bf (a)} Task design. A two-dimensional input (left) is applied to the network for~500 ms to specify a reach direction after which the network must produce output matching the corresponding EMG activity patterns recorded from two arm muscles (right). Each color represents the activity for a specific direction. {\bf (b)} Raster plot showing the activity of a single neuron across all trials (each row) for all conditions (different colors). The Fano factor for this neuron is 1.2. {\bf (c)} Firing rate of the neuron shown in (b). Each color represents the trial-averaged firing rate for a single condition. {\bf (d)} The Fano factor as a function of time computed across all neurons and conditions. {\bf (e)} $\W\s$ for both outputs and all conditions (different colors) on a single trial. {\bf (f)} Firing rates for four network neurons. Some neurons are tuned during input or output periods exclusively (bottom two plots), while most are tuned during both periods (top two plots). {\bf (g)} Trial-averaged firing rate autocorrelation, averaged across all neurons and all conditions.  By the time of movement, the autocorrelation function is near zero, indicating that the tuning between input and movement periods is, on average, uncorrelated.  The time bar in all panels represents 200 ms, and the dot denotes movement onset. Parameters for this example: $N = 5000$, $\mu = -112$, $g_{\fMy} = 33$, $g = 7.2$ mV, $\tilde N = 1000$, $\tilde g = 1.4$.  Also, in this example, the elements of $\utwit$ were chosen randomly and uniformly over the range $\pm 0.76$, and the range for $\Utwit$ was $\pm 0.25$. The refractory period was set to zero in this example. EMGs were filtered with a 4 ms Gaussian window.}
\label{fig:Figure4}	
\end{figure}

Our first example is an autonomous oscillation task that requires the network to generate a self-sustained, temporally complex output~(\hyperref[fig:Figure2]{Figure \ref*{fig:Figure2}}). 
$\Fout$ for this task is a periodic function created by summing sine functions with frequencies of 1, 2, 3, and 5 Hz.  We require that the network generates this output autonomously, therefore~$\Fin = 0$ for this example. Complex, autonomous oscillatory dynamics are a feature of neural circuits involved in repetitive motor acts such locomotion \citep{Marder}.

Initially $\J = 0$, so the activity of the network is determined by the random synaptic input prodvide by~$\J^{\fMy}$, and the neurons exhibit irregular spiking~(\hyperref[fig:Figure2]{Figure \ref*{fig:Figure2}a-c}).  In this initial configuration, the average Fano factor, computed using 100 ms bins, is 0.5, and the average firing rate across the network is 5 Hz. Following the training procedure,  the learned postsynaptic currents $\J\s$ closely match their respective auxiliary target functions~(\hyperref[fig:Figure2]{Figure \ref*{fig:Figure2}d}), and the network output similarly matches the target~$\Fout$~(\hyperref[fig:Figure2]{Figure \ref*{fig:Figure2}e}). Residual chaotic spiking due to~$\J^{\fMy}$~(\hyperref[fig:Figure2]{Figure \ref*{fig:Figure2}c}) and the fact that we are approximating a continuous function by a sum of discontinuous functions cause unavoidable deviations. Nevertheless, a network of 3,000 LIF neurons firing at an average rate of 6.5 Hz with an average Fano factor of 0.25 performs this task with normalized post-training error of 5\% (this error is the variance of the difference between $\W\s$ and $\Fout$ divided by the variance of $\Fout$).

Because the output for this first task can be produced by a linear dynamical system, previous methods could also have been used to construct a functioning spiking network \citep{Eliasmith, BoerlinMachensDeneve}. However, this is no longer true for the following examples.  In addition, it is worth noting that the network we have constructed generates its output as an isolated periodic attractor of a nonlinear dynamical system. The other procedures, in particular that of \citet{BoerlinMachensDeneve}, create networks that reproduce the linear dynamics that generates $\Fout$.  This results in a system that can produce not only $\W\s\approx\Fout$, but also $\W\s\approx\alpha \Fout$ over a continuous range of $\alpha$ values.  This often results in a slow drift in the amplitude of $\W\s$ over time.  The point here is that our procedure solves a different problem than previous procedures, despite the fact that it generates the same output.  The previous procedures were designed to duplicate the linear dynamics that produce $\Fout$, whereas our procedure duplicates $\Fout$ uniquely.

The second task we present is a temporal XOR task that requires the network to categorize the input it receives on a given trial and report this decision through its output.  Each trial for this task consists of a sequence of two pulses appearing as the network input~$\Fin$~(\hyperref[fig:Figure3]{Figure \ref*{fig:Figure3}}).
Each pulse has an amplitude of 0.3, and its duration can be either short (100 ms) or long (300 ms).  The two pulses are separated by 300 ms, and after an additional 300 ms delay the network must respond with either a positive or a negative pulse (with a shape given by 1/2 cycle of a 1 Hz sinewave).  The rule for choosing a positive or negative output is an exclusive OR function of the input sequence (short-short $\rightarrow -$, short-long $\rightarrow +$, long-short $\rightarrow +$, long-long $\rightarrow -$).  The time between trials and the input sequence on each trial are chosen randomly.
 
A network of 3,000 LIF neurons with an average firing rate of~7 Hz can perform this task correctly on 95\% of trials.  As in the autonomous oscillation task, individual neuron spiking activity varies from trial-to-trial due to the effect of~$\J^{\fMy}$. The Fano factor computed across all neurons, all analysis times, and all task conditions is 0.26\@.  This task requires integration of each input pulse, storage of a memory of the first pulse at least until the time of the second pulse, memory of the decision during the delay period before the output is produced, and classification according to the XOR rule.

\subsec{Generating EMG activity during reaching} 

We now turn to an example based on data from an experimental study, with the spiking network generating outputs that match electromyograms (EMGs) recorded in 2 arm muscles of a non-human primate performing a reaching task~\citep{ChurchlandCunningham}.  In this task, a trial begins with the appearance of a target cue at one of eight possible reach directions (task conditions). After a short delay, during which the arm position must be held fixed, a ``go'' cue appears, instructing a reach to  the target. The time between trials and the sequence of reach directions are varied randomly.  

To convey information about target location to the model, we use two network inputs denoted by a two-component vector~$\Fin$ and with amplitudes $2\cos(\theta)$ and $2\sin(\theta)$ where the angle $\theta$ specifies the reach direction~(\hyperref[fig:Figure4]{Figure \ref*{fig:Figure4}a left}). The input is applied for~500 ms and, when it terminates, the network is required to generate two outputs (thus $\Fout$ is also two-dimensional) that match trial-averaged and smoothed EMG recordings from the anterior and posterior deltoid muscles during reaches to the specified target~(\hyperref[fig:Figure4]{Figure \ref*{fig:Figure4}a right \& e}). 

A network of 5,000 neurons with an average firing rate of~6 Hz performs this task with a normalized post-training error of 7\%~(\hyperref[fig:Figure4]{Figure \ref*{fig:Figure4}e}), consistent with another modeling study that used a firing-rate network~\citep{SussilloChurchland}.  The activity of the trained network exhibits several features consistent with recordings from neurons in motor cortex. Individual neurons show a large amount of spiking irregularity that is variable across trials and conditions~(\hyperref[fig:Figure4]{Figure \ref*{fig:Figure4}b}). The Fano factor computed across all neurons and all task conditions drops during task execution~(\hyperref[fig:Figure4]{Figure \ref*{fig:Figure4}d}), consistent with observations across a wide range of cortical areas~\citep{Churchlandetal}. This network shows that EMGs can be generated by a network with a realistic degree of spiking variability.

Another interesting feature of the network activity is the variety of different neural responses. Individual neurons display tuning during the input period, the output period, or across multiple epochs with different tunings~(\hyperref[fig:Figure4]{Figure \ref*{fig:Figure4}f}). As in recordings from motor cortex, consistently tuned neurons represent a minority of the network; most neurons change their tuning during the task~(\hyperref[fig:Figure4]{Figure \ref*{fig:Figure4}g}).  

\begin{figure}
\centerline{\includegraphics[width=4in]{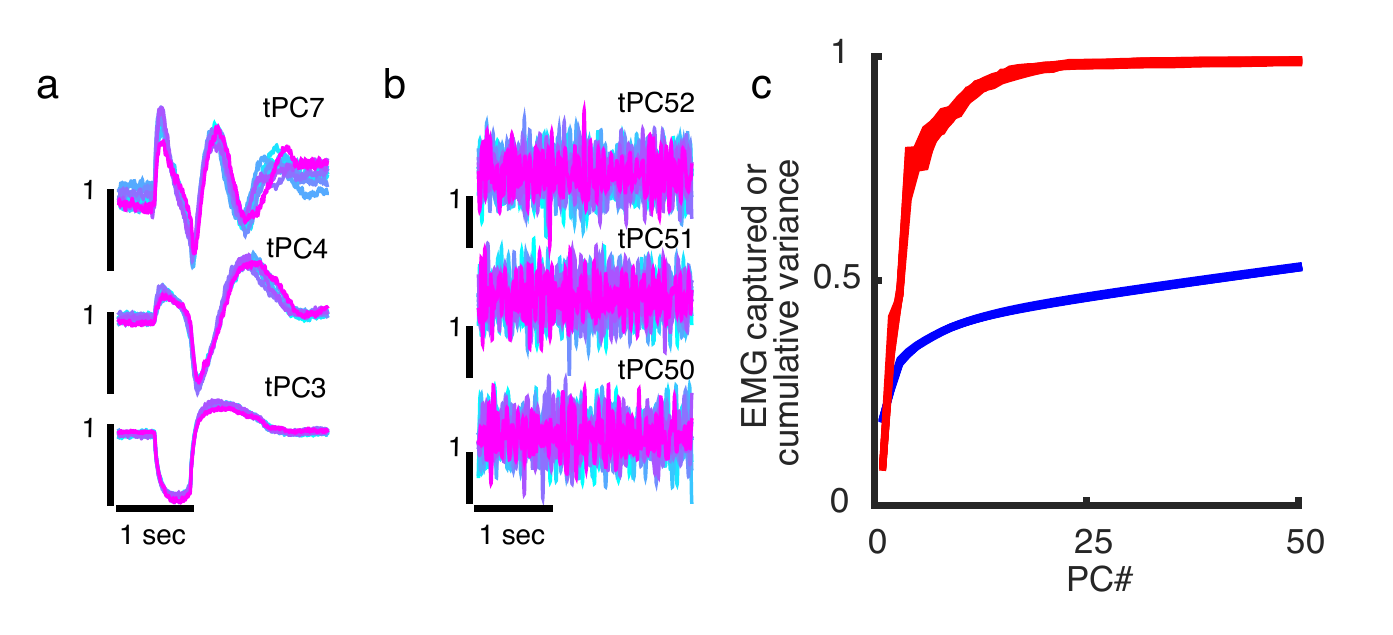}}
\caption{Population level analyses of EMG task activity. {\bf (a)} Temporal PCs 3,4 and 7 computed on a single-trial using data from all neurons and all conditions. Results from three different trials are shown in different colors. {\bf (b)} Same as (a) except displaying temporal PCs 50,51 and 52.  {\bf (c)} Fraction of the total variance captured in successive temporal PCs for single-trial PCA analysis (blue), and fraction of the EMG variance accounted for when regressing against increasing numbers of PCs (red traces corresponding to multiple trials).}
\label{fig:Figure5}	
\end{figure}

To examine the population dynamics of this model, we performed PCA on filtered (4 ms Gaussian filter) single-trial spike trains from a~$T\times NC$ data matrix, where~$T$ is the number of times sampled during a trial (2900),~$N$ is the number of neurons (5000), and~$C$ is the number of reach conditions (8).  We computed the eigenvectors of the~$T \times T$ covariance matrix obtained from these ``data", generating temporal PCs~(\hyperref[fig:Figure5]{Figure \ref*{fig:Figure5}a \& b}).  Each temporal PC represents a temporal function that is strongly represented across the population and across all reach conditions, albeit in different ways across the population on a single trial for each condition.  Two important features emerge from this analysis. First, more prominent single-trial PCs~(\hyperref[fig:Figure5]{Figure \ref*{fig:Figure5}a}) are relatively consistent from trial-to-trial, while less prominent PCs~(\hyperref[fig:Figure5]{Figure \ref*{fig:Figure5}b}) vary more. Second, the prominent PCs fluctuate on a slower time scale than the less prominent PCs. These two features indicate that the more prominent PCs form the basis of the network output on a single trial, while the less prominent PCs reflect network noise. This can be verified by reconstructing the network output using increasing numbers of temporal PCs and calculating the fraction of the output variance captured~(\hyperref[fig:Figure5]{Figure \ref*{fig:Figure5}c red}).  A small number of the leading PCs account for most of the network output variance despite the fact that they account for only a fraction of the full activity variance on a single trial~(\hyperref[fig:Figure5]{Figure \ref*{fig:Figure5}c blue}).

\subsec{Learning constrained connectivities} 

The examples we have presented up to now involve a fully connected $\J$ matrix with no sign constraints.  In other words, no elements were constrained to be zero, and the training procedure could make the sign of any element either positive or negative.  Biological networks tend to be sparse (many elements of $\J$ are fixed at zero) and obey Dale's law, corresponding to excitatory and inhibitory cell types.  This implies that the columns of $\J$ should be labelled either excitatory or inhibitory and constrained to have the appropriate sign ($+$ for excitatory and $-$ for inhibitory). Here we outline a procedure for training a network to solve a task while abiding by these constraints.

In the previous examples, we used a  recursive least squares (RLS) procedure to compute $\J$ because of stability issues that we now explain.  Satisfying equation~\ref{eq:zJsf} accurately assures that $\s$ can be generated self-consistently, but it does not guarantee that the resulting network state will be stable, and if it is unstable the least-squares solution is useless.  We find that the use of RLS resolves this problem.  As explained previously \citep{SussilloAbbott}, RLS provides stability because the fluctuations produced by the network when it is doing the task are sampled during the recursive procedure, allowing adjustments to be made that quench instabilities.  However, when sparseness and especially sign constraints are impossed, use of RLS becomes impractical.  Instead we must resort to a batch least-squares (BLS) algorithm.  

The BLS algorithm computes $\J$ on the basis of samples of $\s$ that must be provided.  To assure a stable solution, these samples should not only characterize the activity of a network doing the task, they should include the typical fluctuations that arise during network operation.  To obtain such samples, we perform the network training in two steps.  First, we train a fully connected and sign-unconstrained network using the RLS procedure, just as in the previous examples.  We sample $\s$ from this network during the training process, and use these samples to solve the BLS problem while enforcing the desired constraints. 

\begin{figure}
\centerline{\includegraphics[width=4in]{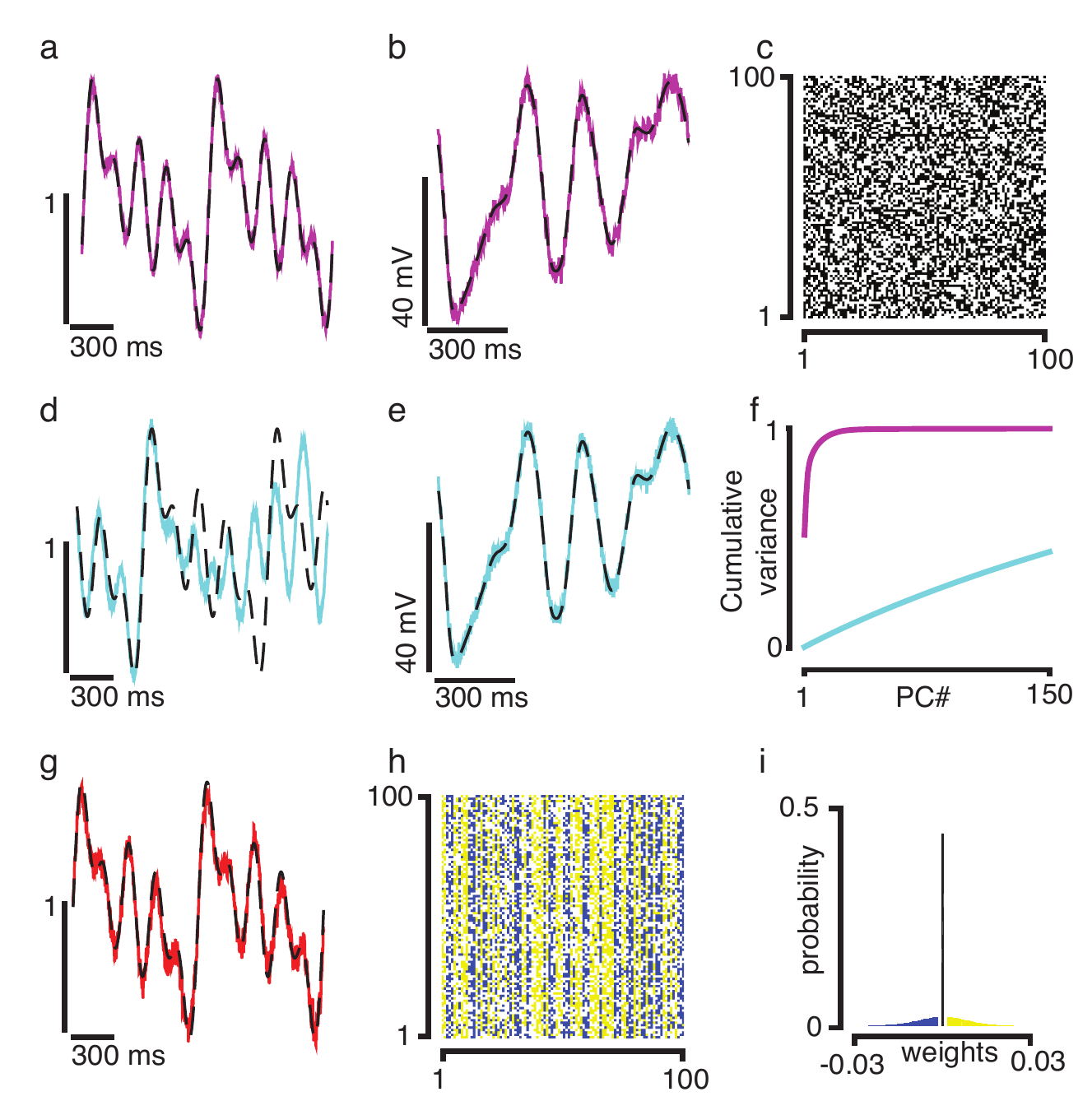}}
\caption{Performing the autonomous oscillation task (Figure~\ref{fig:Figure2}) with constrained connectivity.  {\bf (a)} $\Fout$ (dashed black) and $\W\s$ (magenta) from a network with 50\% sparse connectivity. {\bf (b)} $\FJ$ (dashed black) and $\J\s$ (magenta) for one neuron during training. Residual fluctuations can be seen, critical for stability after learning.  {\bf (c)} Entries of~$\J$ for 100 neurons. {\bf (d)} Same as (a) except residuals from RLS solutions were shuffled before BLS was performed.  {\bf (e)} Same as (b), except showing shuffled residuals. {\bf (f)} Cumulative variance of successive PCs of the spatial covariance matrix of the RLS residuals (magenta) and the shuffled residuals (cyan). {\bf (g)} $\Fout$ (dashed black) and $\W\s$ (red) from a network with 50\% sparseness satisfying Dale's Law (with 50\% excitatory and $50\%$ inhibitory neurons). {\bf (h)} Entries of~$\J$ for (g) for 100 neurons. {\bf (i)} Histogram of the entries of~$\J$ for (g). Network parameters: $N = 1000$, $\mu = g_{\fMy} = 0$, $g =  15$ mV\@.}
\label{fig:Figure6}	
\end{figure}

Applying this two-step procedure, we can construct networks that perform the autonomous oscillation task of Figure~\ref{fig:Figure2} with 50\% sparseness, either without~(\hyperref[fig:Figure6]{Figure \ref*{fig:Figure6}a-c}) or with~(\hyperref[fig:Figure6]{Figure \ref*{fig:Figure6}g-i}) a Dale's law constraint. The normalized post-training error for both networks on this task is 5\%, although to obtain this level we had to allow an average firing rate of~22 Hz.  In addition, the random connectivity represented by $\J^{\fMy}$ was not included in this case because, for the number of neurons we used (1000), these networks are somewhat fragile to chaotic fluctuations.  This fragility could be reduced by using more neurons, but even with BLS the computations, especially for the sign-constrained case, are quite lengthy.

We argued that the two-step training procedure was needed to sample network fluctuations properly.  Could we have simply added white noise to samples of $\s$ that did not contain the actual fluctuations produced by an operating network~\citep{Jaeger, Eliasmith}?  PCA on the covariance matrix of the network fluctuations obtained during RLS training shows that most of their variance is captured by a small number of PCs~(\hyperref[fig:Figure6]{Figure \ref*{fig:Figure6}f magenta}), indicating significant spatial correlations.  The temporal autocorrelation function for residual errors also showed significant correlation (not shown).  To understand the role of these correlations, we created a  dataset of $\s$ with the actual network fluctuations replaced by shuffled fluctuations.  Although the shuffled synaptic input~(\hyperref[fig:Figure6]{Figure \ref*{fig:Figure6}e}) is very similar to the unshuffled input~(\hyperref[fig:Figure6]{Figure \ref*{fig:Figure6}b}), use of the shuffled data set resulted in poor performance of the trained network~(\hyperref[fig:Figure6]{Figure \ref*{fig:Figure6}d}).  This is because the shuffled data fail to capture the correlations present in the actual network fluctuations~(\hyperref[fig:Figure6]{Figure \ref*{fig:Figure6}f cyan}). These results reaffirm the conclusion that the RLS algorithm is effective for sampling network instabilities, and that the fluctuations obtained in this way can  be used effectively to obtain constrained BLS values for $\J$.

\sec{Discussion}

We have developed a framework for constructing recurrent spiking neural networks that perform the types of tasks solved by biological neural circuits and that can be made compatible with biophysical constraints on connectivity.  In this approach, the first step in producing a spiking system that implements a task is to identify a continuous-variable dynamical system that can, at least in principle, perform the task. Previous approaches to building models that perform tasks also resorted to identifying continuous analog systems.  A key distinction, however, is that by exploiting the rich dynamics of externally driven, randomly connected, continuous-variable models, we can apply our approach to cases where a dynamic description of the task is not readily apparent.  In general, any continuous variable network that can implement a task should generate useful auxiliary targets for training a spiking model.  An intriguing future direction would be to use continuous-variable networks trained with back-propagation \citep{MartensSutskever, SussilloChurchland, SussilloRev} for this purpose.  Another recent proposal for training spiking networks also makes use of continuous-variable network dynamics, but in this interesting approach, a spiking network is constructed to duplicate the dynamics of a continuous-variable model, and then it is trained essentially as if it were the continuous model \citep{Thalmeieretal}.

Our work involves a more complex map from the continuous variables of a ``rate" model to the action potentials of a spiking model.  The simplest map of this type assigns a population of spiking neurons to each unit of the continuous-variable model such that their collective activity represents the continuous ``firing-rate".  If we had followed this path, our spiking networks would have likely involved hundreds of thousands to millions of model neurons.  In our approach, only a few times as many spiking neurons as continuous variable units are needed.  Performing PCA on the activity of a continuous-variable network reveals that relatively small number of PCs capture a large fraction of the variance \citep{RajanAbbottSompolinsky, SussilloAbbott}.  Thus, the unit activity in these networks is redundant, and matching every unit with a different population of spiking neurons is wasteful because this amounts to representing the same PCs over and over again.  Our approach avoids this problem by distributing continuous signals from the entire continuous-variable network to overlapping pools of spiking neurons. 
 
Our work also involves a novel interpretation of continuous variable models and the outputs of their units. These models are typically considered to be approximations of spiking models \citep{Ermentrout, Gerstner, Shriki, Ostojic}.  We do not interpret the ``firing rates" of units in continuous-variable networks as measures of the spiking rates of any neuron or collection of neurons in our spiking networks (in fact, these ``firing rates" can be negative, which is why we avoid the term firing-rate network).  Instead, the continuous-variable networks are generators of the principle components upon which the dynamics of both networks are based. PCA applied to $\s$ of the spiking network and to  $H(\x)$ of the continuous-variable network yield very similar results, at least for the predominant PCs.  For example, the leading 10 temporal PCs account for more than 90\% of the total variance for both networks, and the median of the principle angles between the subspaces defined by these two sets of 10 vectors is 5 degrees.  This indicates that the temporal signals that dominate the dynamics and output of these two different types of networks are extremely similar. 

We extracted a new set of auxiliary targets by projecting the original set onto a relatively small number of PCs of the continuous-variable network, and this did not have a detrimental impact on network training.  We did this for the autonomous oscillator task using just 12 PCs, and the normalized error for the resulting network was similar to the error for the network shown in Figure~\ref{fig:Figure2}. This confirms that the key information being passed from the continuous-variable network to the spiking network is carried by the leading PCs.  The continuous-variable network is used in our procedure as a way of computing the PCs relevant to a task.  If these can be obtained in another way, the spiking network could be trained directly from the PCs.  One way of doing this is to extract the PCs directly from data, as has been done in other studies \citep{Fisheretal, RajanHarveyTank}. 

Finally, our approach strongly supports the use of continuous-variable models to analyze and understand neural circuits.  However, it is important to appreciate that the connection between spiking and continuous-variable networks is subtle.  In our procedure, the connectivity and nonlinearity in the continuous network bear no relation to the corresponding features of the spiking model, and the continuous network is not unique.  Furthermore, the signals that allow a task to be performed are only apparent at the population level.  In addition, since our spiking networks are not constructed by a rational design process, it may not be immediately apparent how they work.  Nevertheless, the underlying continuous-variable model, and especially its leading PCs, capture the essence of how the spiking network operates and tools exist for understanding this operation in detail \citep{SussilloBarak}.  These models and methods should do the same for experimental data. 

\newpage
\sec{Acknowledgments}

We are grateful to Antonio Lara for providing the EMG data.  Research supported by NIH grant MH093338 and by the Simons Collaboration for the Global Brain, the Gatsby Charitable Foundation, the Swartz Foundation, the Mathers Foundation and the Kavli Institute for Brain Science at Columbia University.  B.D. was supported by a National Science Foundation Graduate Research Fellowship.

\end{document}